\documentclass[default,iicol]{sn-jnl}

\jyear{2021}%

\usepackage{soul,xcolor}
\begin{document}
\setstcolor{red}
\title[ ]{Charge Migration in Heterocyclic Five-Membered Rings}


\author[1]{\fnm{Sucharita} \sur{Giri}}

\author[1]{\fnm{Gopal} \sur{Dixit}}\email{gdixit@phy.iitb.ac.in}

\author*[2]{\fnm{Jean Christophe} \sur{Tremblay}}\email{tremblay1@univ-lorraine.fr}

\affil[1]{\orgdiv{Department of Physics}, \orgname{Indian Institute of Technology Bombay}, \orgaddress{\street{Powai}, \city{Mumbai}, \postcode{400076}, \country{India}}}

\affil[2]{\orgname{CNRS-Universit\'e de Lorraine, UMR 7019, }, \orgaddress{\street{1 Bd Arago}, \city{Metz}, \postcode{57070}, \country{France}}}


\abstract{This contribution presents numerical simulations of $N$-electron dynamics in heterocyclic five-membered ring molecules
to shed light on the effect of molecular symmetry on charge migration.
Laser-driven dynamics is studied using the hybrid time-dependent density functional theory/configuration methodology,
and the ensuing field-free charge migration is investigated by means of transient electronic flux density maps.
Our results demonstrate that the charge migration in aromatic rings is sensitive to the presence of {hetero}atoms such as oxygen and nitrogen.
Their presence within the ring induces significant modifications of the character in the ground and low-lying electronic states,
which is imprinted in the charge migration mechanism.
}




\maketitle
\section{Introduction}\label{sec1}

Recent developments in laser technologies facilitate  the generation of ultrashort and intense laser pulses. 
The availability of such laser pulses  allows studying ultrafast  processes in molecules on their natural timescale~\cite{krausz2009attosecond}.  
During laser-molecule interaction, electrons are the first to respond to the action of an ultrashort laser 
pulse. 
As a result of the interaction, a superposition of several electronic states in the form of 
an electronic wave packet is created, which triggers  an electronic charge distribution into motion. 
The migration of  electronic charge around the atoms of the molecular scaffold 
affects the outcomes of chemical reactions and biological functions~\cite{cederbaum1999ultrafast,  kuleff2005multielectron, breidbach2003migration, remacle2006electronic, breidbach2005universal, nisoli2017attosecond, bredtmann2014x}. 
Thus, inducing and controlling the charge migration in molecules  have emerged as one of the cornerstone topics of attosecond science in recent years~\cite{folorunso2021molecular, kraus2015measurement, calegari2014ultrafast, he2022filming}.  

Charge migration is connected  to the transiently-evolving electronic flux  density (EFD) 
as a result of the quantum version of the continuity equation~\cite{sakurai2006advanced}. 
The EFD is a vector field, which provides mechanistic information about the  spatial distribution  of the charge migration along the direction of flow.  
Over the years, EFD has emerged as an important tool to gain detailed insights {into} the underlying mechanism of ultrafast charge migration in systems ranging from  molecules to nano-materials~\cite{nagashima2009electron, giri2020time, okuyama2009electron, giri2021imaging, giri2022probing, shao2020gal, hermann2016ultrafast, pohl2019imaging, shao2021bio, sobottka2020tuning}.  

Charge migration of valence electrons and corresponding EFD in ring-shaped molecules have received ample 
attention in the last few years. 
Charge migration in magnesium porphyrin has been   
studied by means of EFD~\cite{nam2020monitoring, tremblay2021time, barth2006periodic, barth2006unidirectional, koksal2017effect}.
Control over aromaticity in benzene is understood by  analysing charge migration ~\cite{ulusoy2011correlated}. 
Multidirectional angular EFD with a pincer-type flow during charge migration in benzene has been observed,
which further highlights that a significant fraction of electrons is inactive during the flow. 
Furthermore, it has been found that the nature and direction of EFD can be controlled by selective excitation using optimised laser pulses~\cite{hermann2016multidirectional, jia2017quantum, hermann2017attosecond, hermann2020probing}. 
At this juncture, it is natural to ponder how charge migration and the corresponding EFD will differ
if  one reduces the symmetry of the molecule by transiting from six-membered benzene to five-membered 
ring-shaped molecules. This is the main aim of the present work.  

Both cyclopentene and cyclopentadiene are homocyclic and nonaromatic  pentagonal molecules of carbon atoms. 
Moreover, both molecules are ionic  in nature, which makes them less attractive for the present  study.  
On the other hand, pyrrole, furan, and oxazole are five-membered ring-shaped  neutral molecules. 
Thus, these three molecules appear suitable to serve our purpose to explore the role of symmetry reduction on charge migration and EFD. 
These three heterocyclic aromatic molecules are present in many important biological systems as a fundamental unit and have enormous applications~\cite{demingos2021first}. 
As such, the static electronic properties of pyrrole, furan, and oxazole have received significant attention from theoreticians~\cite{christiansen1999electronic, roos2002theoretical, lee1996molecular, serrano1993theoretical, wan2000electronic, burcl2002study, palmer1995electronic, christiansen1998electronic}.
Further, a few non-equilibrium dynamics in three molecules have been documented in Refs.~\cite{cao2016excited, geng2020time, carrascosa2021mapping, barbatti2010non, barbatti2006nonadiabatic, yong2021ultrafast}.

The present work provides a detailed and comparative study of charge migration of valence electrons in 
pyrrole, furan, and oxazole. 
In the following, we demonstrate how the charge migration is sensitive to the symmetry reduction, i.e., progressing 
from benzene with $\mathcal{D}_{\textrm{6h}}$ symmetry to these five-membered ring-shaped  
molecules with $\mathcal{C}_{\textrm{2v}}$ and $\mathcal{C}_{\textrm{s}}$ symmetries. 
Furthermore, we will discuss how the nature of the electronic charge  and flux densities  during charge migration
is altered by the presence of {hetero}atoms, i.e.,  nitrogen in pyrrole, oxygen in furan; and nitrogen and oxygen in oxazole.   

\section{Computational Methods}\label{sec2}
Charge migration is simulated by solving numerically the time-dependent Schr\"odinger equation 
\begin{equation}\label{eq.tdse}
i\hbar\frac{\partial}{\partial t}\vert\Psi (t) \rangle = \mathcal{H}(t)\vert\Psi(t) \rangle,
\end{equation}
where $\mathcal{H}(t)$ is the time-dependent Hamiltonian describing an $N$-electron system interacting with 
an ultrashort laser pulse, and treated in the semi-classical dipole approximation as
\begin{equation}\label{eq.hamiltonian}
 \mathcal{H}(t) = \mathcal{H}_{0} - \hat{\mu} \cdot \mathbf{F}(t). 
\end{equation}
Here, $\mathcal{H}_{0}$ is the field-free molecular Hamiltonian within the clamped nuclei approximation, 
$\hat{\mu}$ is the dipole moment operator, and $\mathbf{F}(t)$ is the electric field of the incident laser. 
An electronic wave-packet is created in a molecule upon application of an ultrashort laser.
Within the hybrid time-dependent density functional theory-configuration interaction (TDDFT/CI) method, 
the $N$-electron wave-packet is expressed as a linear combination of many-body excited electronic states
\begin{equation}\label{eq.wavefunction}
    \Psi (\mathbf{r}^N, t) = \sum_{k = 0}^{N_{\textrm{states}}} C_{k} (t) \Phi_{k} (\mathbf{r}^N),
\end{equation}
with expansion coefficients $C_{k} (t)$.
$N_{\textrm{states}}$ is the number of many-electron states, $\Phi_{k} (\mathbf{r}^N)$, necessary to describe the wave-packet and its response in the presence of an ultrashort, intense laser pulse.
In the CI Singles picture, each $N$-electron excited state is written as a linear combination of singly-excited configuration state functions (CSFs) as
\begin{equation}\label{eq.tdci}
\Phi_{k} (\mathbf{r}^N) = D_{0, k} \Phi_{0} (\mathbf{r}^N) + \sum_{ar} D_{a, k}^{r} \Phi_{a}^{r} (\mathbf{r}^N). 
\end{equation}
The first term on the right-hand side represents the ground state Slater determinant $\Phi_{0} (\mathbf{r}^N)$, obtained from Kohn-Sham density functional theory.
The second term contains CSFs describing single excitations from an occupied molecular orbital $``a''$ to a virtual molecular orbital $``r''$. 
In TDDFT/CI, the expansion coefficients  $D_{0, k}$ and  $D_{a, k}^{r}$ are obtained using linear-response
TDDFT in the Tamm-Dancoff approximation.
This combination was shown to provide good energetics while keeping the computational cost low \cite{klinkusch2016}.

To analyse the laser-induced charge migration, the time-evolution of the one-electron density $\rho(\mathbf{r}, t)$
and the associated EFD $j(\mathbf{r}, t)$ are reconstructed from the $N$-electron wavepacket dynamics.
These two observable quantities are connected via the quantum continuity equation as
\begin{equation}\label{eq.continuity}
    \frac{\partial}{\partial t} \rho(\mathbf{r}, t) = -\vec{\nabla} \cdot \mathbf{j} (\mathbf{r}, t).
\end{equation}
Here, $\rho(\mathbf{r}, t)$ is the expectation value of the one-electron density operator as
\begin{eqnarray}\label{rho_op}
\hat{\rho} \left( \mathbf{r}\right) = \sum_{k=1}^{N} \delta\left( \mathbf{r} - \mathbf{r}_{k} \right).
\end{eqnarray}
The latter also gives access to the electron density in a given state,
$\rho_m(\mathbf{r})=\langle \Phi_{m} \vert \hat{\rho} \left( \mathbf{r}\right)\vert \Phi_{m} \rangle$,
and the transition densities $\rho_{mn}=\langle \Phi_{m}\vert \hat{\rho} \left( \mathbf{r}\right)\vert \Phi_{n}\rangle$.
Similarly, the EFD can be written in the following operator form 
\begin{eqnarray}\label{j_op}
\hat{j} \left( \mathbf{r}\right) = \frac{1}{2}\sum_{k=1}^{N} \left(\delta(\mathbf{r}-\mathbf{r}_k)\hat{p}_{k} + \hat{p}^{\dagger}_{k}\delta(\mathbf{r}-\mathbf{r}_k)\right), 
\end{eqnarray}
where $\hat{p}_{k}=-i\hbar\vec{\nabla}_{k}$ is the momentum operator of electron $k$. 
The transient flux density maps are then the expectation values of this operator. 

To achieve the convergence of the laser-induced charge migration simulations, 
the $N_\textrm{states} = $ 24, 22, and 32  lowest-lying bound excited states are used for pyrrole, furan and oxazole, respectively. 
The CAM-B3LYP functional and aug-cc-pVQZ basis sets on all atoms are used to compute these electronic states~\cite{04:YTH:cam, dunning1989ccpvxz}.
All electronic structure calculations are {performed} with the Gaussian16 program \cite{frisch2016gaussian} and the open-source toolbox detCI@ORBKIT is used 
post-process the information obtained from the quantum chemistry package and to calculate all the required observable~\cite{hermann2016orbkit, pohl2017open, hermann2017open}. The dynamical simulations are performed using an in-house implementation of the TDCI method \cite{08:TKS:rhoTDCI,10:TKKS:resp,11:TKKS:irhoTDCI}.

{The time evolution of the many-electron population dynamics in the three molecules is shown for a  20\,fs sine-squared pulse linearly polarised along the $y$-axis.
The carrier frequency of 6.26 eV, 6.26 eV, and 6.37 eV and peak intensity of 3.7$\times$10$^{13}$ W/cm$^{2}$, 2.8$\times$10$^{13}$ W/cm$^{2}$, and 7$\times$10$^{13}$ W/cm$^{2}$ pulses are used for inducing charge migration in pyrrole, furan, and oxazole, respectively.}
\section{Results and Discussion}
\begin{figure}[b!]
\includegraphics[width = \linewidth]{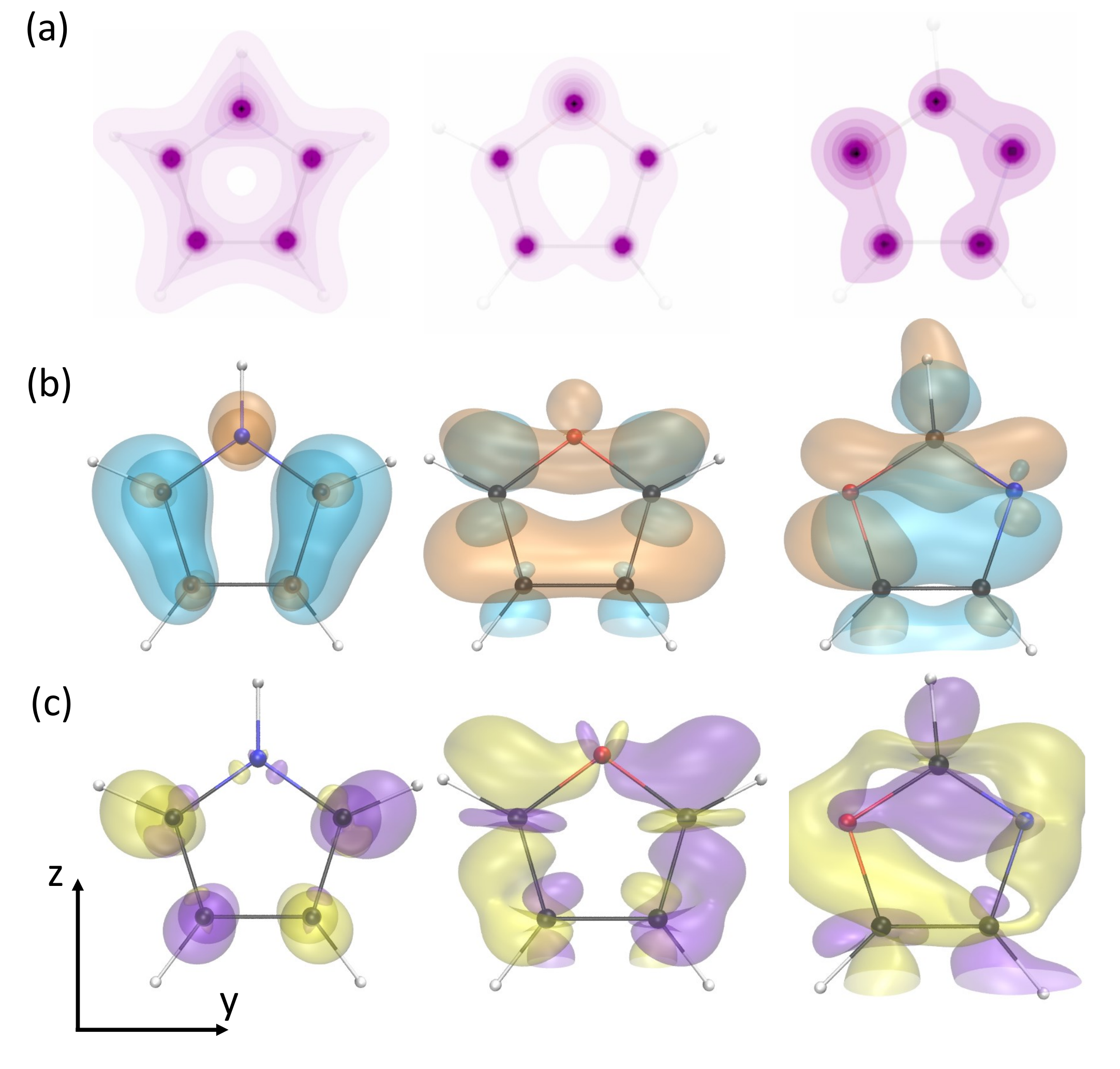}
\caption{Depiction of the electronic structure in the five-membered ring molecules. 
{5$^{\text{th}}$, 2$^{\text{nd}}$, and 1$^{\text{st}}$ excited states of pyrrole, furan, and oxazole, respectively are represented from left to right.} 
(a) Ground state electron density (isocontours equally separated by 0.001\,a.u.).
(b) Natural transition orbitals. Orange (cyan) surfaces represent the particle (hole) densities, respectively {for isosurface value of 0.005}. 
(c) Difference electron densities between target excited and ground electronic states, $\rho_{m}(\mathbf{r})-\rho_{0}(\mathbf{r})$. {Violet(yellow)} colour represents regions of density increase (depletion) {for isosurface value of +0.005 (-0.005)}.}
\label{fig1}
\end{figure}
To compare the charge migration in pyrrole, furan, and oxazole on an equal footing, {a few} conditions are considered.
The first one is a static consideration, where the target excited states are chosen to have a similar character.
This can be achieved by exciting states that, {\it e.g.}, create nodal structures around the heteroatoms. 
{Further, in order to populate a target excited state, it must be optically accessible.
That is, the intensity of the transition from the ground to target excited state (proportional to the transition dipole moment $\mu^{2}$) must be high.
From a dynamical perspective, charge migration should be faster than the timescale of nuclear rearrangement,
ideally in the sub-femtosecond regime, yet accessible with a few-femtosecond  laser pulse.
The target excited states satisfying both constraints will thus have similar energies and transition dipoles.
Further, understanding the dynamics of the electronic density and EFD after such a pump pulse is 
greatly simplified if  the wavepacket is composed of only a few components in the excited state manifold.
This is achieved by exciting only a small fraction of electronic  populations 
and optimizing the laser parameters (field intensity and carrier frequency),
thereby leading to a controlled degree of charge migration.}

\begin{figure*}[tb!]
\includegraphics[width = \linewidth]{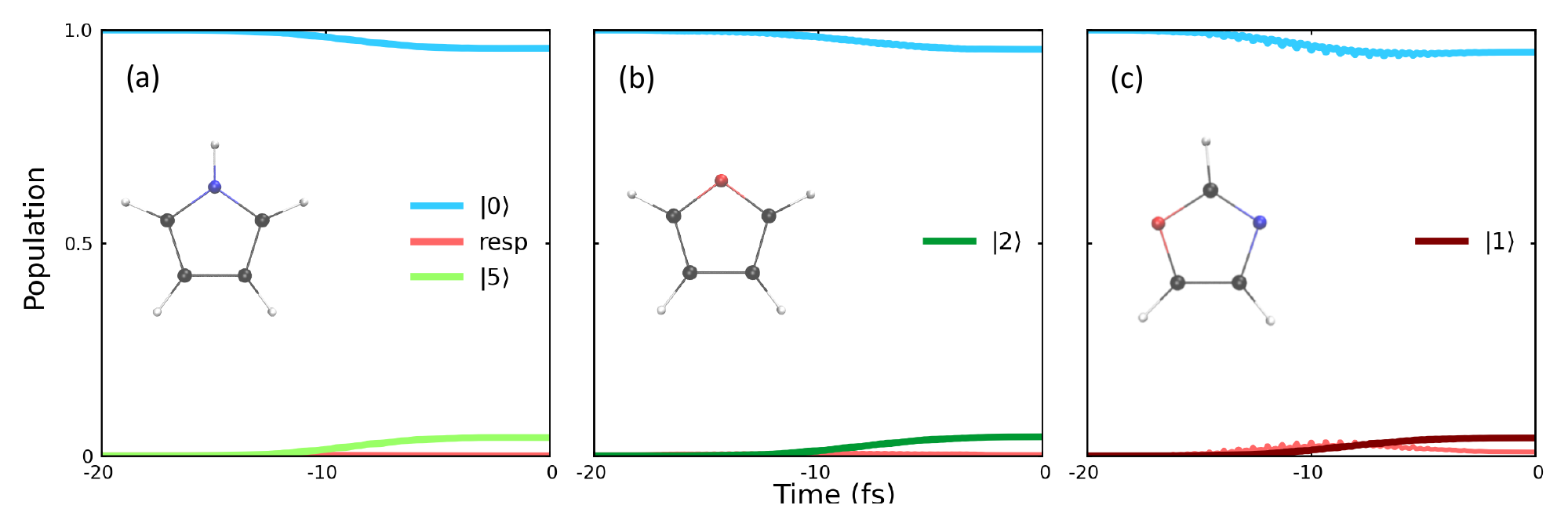}
\caption{Laser-induced population dynamics  in heterocyclic five-membered ring molecules
triggered by 20\,fs sine-squared pulse linearly polarised along $y$-axis.
The label ``resp'' indicates  the residual population of all other electronic states, which acts as a non-linear response during the excitation.
The three molecules are shown in ball-and-stick representation, which are projected in the $yz$-plane throughout this work.
White, grey, blue, and red colours represent hydrogen, carbon, nitrogen and oxygen, respectively.
(a) Pyrrole: carrier frequency resonant with the energy of the 5$^{\text{th}}$ electronic state (6.26 eV); peak intensity of 3.7$\times$10$^{13}$ W/cm$^{2}$.
(b) Furan: carrier frequency resonant with the energy of the 2$^{\text{nd}}$ electronic state (6.26 eV); peak intensity of 2.8$\times$10$^{13}$ W/cm$^{2}$.
(c) Oxazole: carrier frequency resonant with the energy of the 1$^{\text{st}}$ electronic state (6.37 eV); peak intensity of 7$\times$10$^{13}$ W/cm$^{2}$.}
\label{population}
\end{figure*}

The charge distribution corresponding to the ground electronic state of the three molecules is presented in Fig.~\ref{fig1}(a). 
All three molecules are projected in the $yz$-plane throughout this work. 
Pyrrole and furan are symmetric with respect to reflection about the $z$-axis that bisects molecule through the heteroatom and the opposing CC bond, 
and the ground state densities both transform according to the totally symmetric point IRREP of the $\mathcal{C}_{\textrm{2v}}$ point group.
The presence of the nitrogen atom (blue) appears to favour the delocalisation of the electrons over the whole ring in the $\pi$-orbitals.
The more strongly electronegative oxygen atom (red) leads to a more compact ground state density in furan (central panel), which is more concentrated 
at the heteroatom.
Due to the presence of both oxygen and nitrogen atoms on the two sides of the pentagon, oxazole is not symmetric with respect to a reflection in the plane
and retains only the operations of the $\mathcal{C}_{\textrm{s}}$ point group. 

To identify interesting target state in each molecule, the natural transition orbitals (NTOs) of {\color{blue} all states up to the ionisation threshold have been inspected}.
The NTOs of the lowest excited state accessible via a linearly polarised field along the $y$ axis are shown in Fig.~\ref{fig1}(b).
These correspond to the fifth, second, and first excited states in pyrrole, furan, and oxazole, respectively.
{The transition intensities of these target excited states are much higher than the transition intensities
from the ground to all other excited states of the respective molecule.}
NTOs offer a compact depiction of electronic transitions from a reference state in terms {of} particle and hole densities~\cite{martin2003natural}. 
In all three molecules, the excitation indeed creates nodal structure around the heteroatoms but the excitation characters are radically different.
The hole density ({cyan}) in pyrrole is found to be delocalised on both sides of the ring on the carbon atoms, with a clear $\pi$ character and a node along the $y$-axis.
The particle density (orange) is more strongly localised close to the atoms of the ring, with a $\pi$ character around the nitrogen atom.
The structure is very different in furan, where the hole density is localised around the carbon atoms and exhibits a stronger {in-plane bonding character (reminiscent of a $\sigma$ character, in opposition to an out-of-plane $\pi$-character).} 
The same character is found in the particle density, which is delocalised along the C-O-C fragment, as well as through space between the two C=C bonds.
Both excitations in pyrrole and furan retain the structure of the $\mathcal{C}_{\textrm{2v}}$ point group.
The NTO densities in oxazole are more similar to the furan case, although it is the hole density that has the through-space structure.

\begin{figure*}[t!]
\includegraphics[width =  \linewidth]{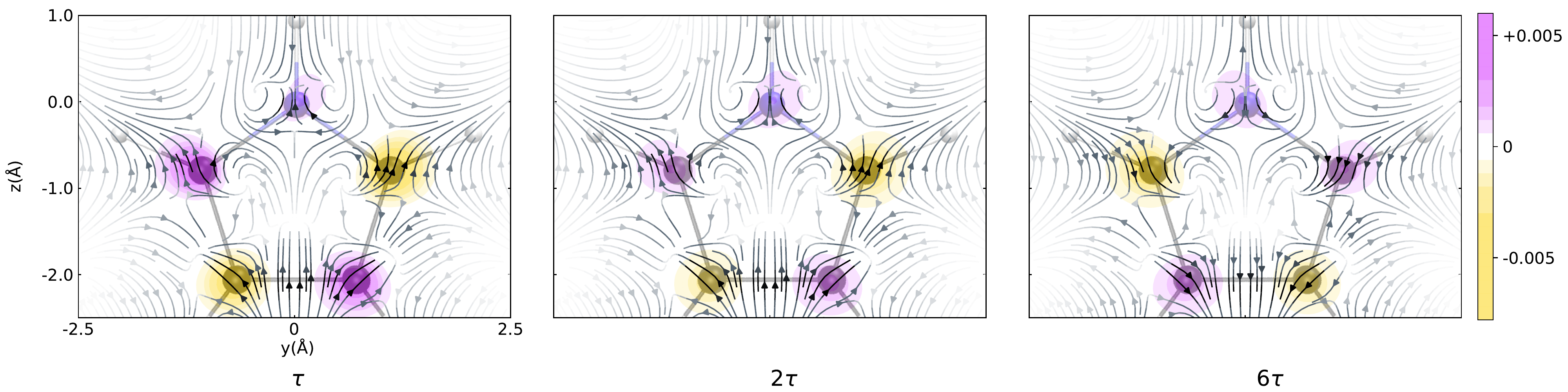}
\caption{Electronic flux density and time-dependent difference electronic density  associated 
with charge migration in pyrrole at $\tau$, 2$\tau$, and 6$\tau$.
Here, $\tau = \textrm{T}/8$ with  
$\textrm{T} =  660$ attoseconds  as  the characteristic timescale of the charge migration in pyrrole.
Ground-state electron density is subtracted  {from} the time-dependent difference electronic density at each subsequent time-step. 
The colour of the streamline arrows varies from white to black according to their increasing intensity. 
The {violet (yellow)} contours present the enhancement (depletion)  of the electronic density with respect to the ground state density. 
For better visualisation, the projection of pyrrole in the $yz$-plane is superposed in ball-stick representation.}
\label{pyrrole_stream}
\end{figure*}

The difference densities between the ground and target states are also shown in Fig.~\ref{fig1}(c).
The strong $\pi$-{type characters, {\it i.e.}, the out-of-plane bonding characters of the excitation in pyrrole is noticed.
In opposition to $\pi$ character, a clear nodal structure in the plane of the molecule analogous to} the $\sigma$ character of the excitations in furan and oxazole
can be readily recognised. More importantly, the difference densities reveal the symmetry of the excited states,
which are found to transform according to the B\textsubscript{2} IRREP of the $\mathcal{C}_{\textrm{2v}}$ point group
for pyrrole and furan. These important differences should lead to markedly different charge migration dynamics after laser excitation.

We  employ a linearly {polarised} pulse along the $y$ direction to induce the charge migration.
The resultant population dynamics are presented in Fig.~\ref{population}.
The pulses for the three molecules have comparable peak intensities, with similar carrier frequencies tuned at the respective target transitions.
The peak intensities of the pulses, in excess of 10$^{13}$W/cm$^2$, are tuned to obtain a similar population in the target states,
hence leading to a similar contrast in the charge migration patterns. As evident from the figure, the target excited states reach around $4 \%$ population in all three cases. {The values of the transition dipole ($\mu^{2}$) to the target excited states
are 0.9898 $ea_0$, 1.1071 $ea_0$, and 1.0048 $ea_0$ for pyrrole (fifth), furan (second), and oxygen (first), respectively.
This compares favourably to all other transitions with values lower than 0.25 $ea_0$ in all three molecules.}

The population in all off-resonant electronic states are termed as ``resp'' in the figure, as they mediate the non-linear response of the 
electronic density to the strong external electric field.
The electronic response is found to be more important in oxazole than in either pyrrole or furan.
{During} the pulse, the population of these non-resonant states is exactly zero for pyrrole and furan, {but non-zero} in oxazole. {This is because of the strong transition dipole moments to other excited states in oxazole,
which stems from its  different chemical nature compared to the other two molecules.}
The field-free charge migration sets into motion at time zero, after the pulse. 
Since the target excited states in the three systems have similar excitation energies with respect to the ground electronic state,
the charge migration will have a similar oscillation period around 650\,as.

To shed light on the effect of symmetry on field-free charge migration on its characteristic timescale,
we analyse snapshots of the EFD and of the associated time-dependent difference electronic density.
For the latter, the contribution from the electronic ground state is subtracted from the wave packet at each time-step. 

\begin{figure*}[t!]
\includegraphics[width = \linewidth]{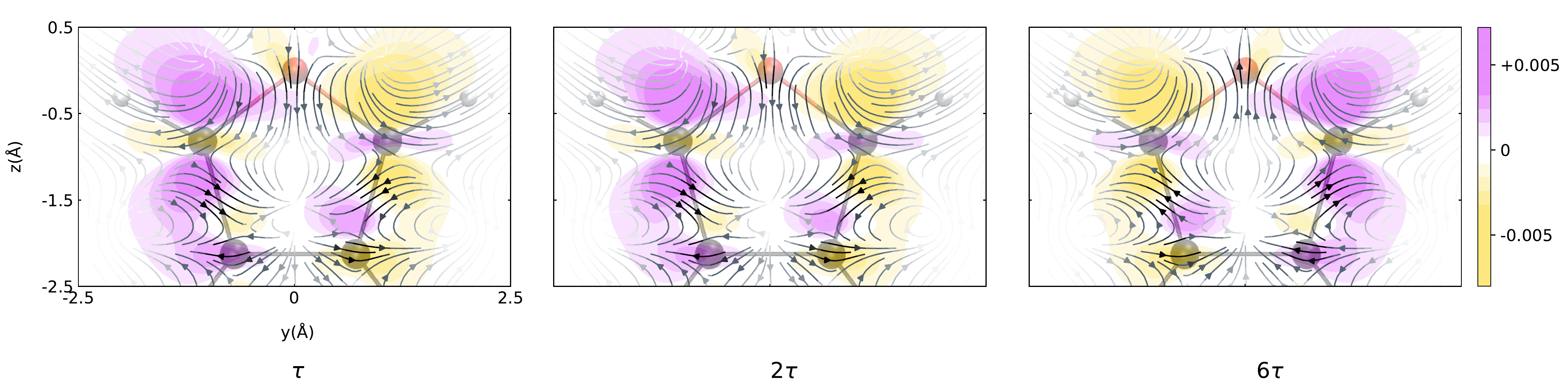}
\caption{Same as Fig.~\ref{pyrrole_stream} for furan  with $\textrm{T} =  660$ attoseconds  as  the characteristic timescale.} 
\label{furan_stream}
\end{figure*}

Figure~\ref{pyrrole_stream} presents the EFD and the difference density at three time steps for pyrrole. 
The overall behaviour of the difference densities at all times reflects that they are mostly located  
around the atoms, which indicates charge localization.
This corresponds well to the picture offered by the density difference in the left panel of Fig.~\ref{fig1}(c).
{The strong charge localisation around the atoms should not affect strongly the bonding properties in pyrrole.
Consequently, it is unlikely that the excitation chosen would lead to any bond breaking, ring-opening or other photochemical reaction.}
Moreover, the difference densities  are positive and negative in alternate carbon atoms, whereas  
nitrogen is always surrounded by positive electron density. 
At $\tau$ and 2$\tau$, the positive contour around  nitrogen  is tilted towards the positive $y$-axis
as evident from the figure.  This situation is reversed at 6$\tau$. 
The difference densities around carbon atoms are anti-symmetric about $y$-axis at all times. 
The overall intensity of the difference density reduces from $\tau $ to  $2\tau$. 
There is a phase shift  of ${\pi}/{2}$ from  $2\tau$ to $6\tau$, which is in accordance with the 
half time-period  of the charge migration as evident from the figure. 
The characteristic timescale form charge migration in pyrrole is $\textrm{T} =  660$ attoseconds
for the specific superposition state created here. 

The electron flow can be understood better in terms of the EFD. 
As evident from Fig.~\ref{pyrrole_stream}, the direction of the electron flow is mostly around the atoms. 
The streamlines start from the atomic positions and terminate around the bonds at all times.
This reflects the atom-to-bond charge migration in pyrrole, which corresponds well to the picture offered by the NTO in Fig.~\ref{fig1}(b).
Indeed, the particle density is observed to be strongly localised around the atoms.
Consequently, the streamlines appear as piecewise disjoint lines of flow around each atom.  
From a symmetry point of view, there is no general electron flow from left to right or from upward to downward.
At $\tau$, the flow around the bottom two carbon atoms is upward inside and downward outside of the pyrrole ring. 
Similar to the difference densities, the streamlines follow the same phase reversal from $2\tau$ to $6\tau$. 
The EFD transforms according to the totally symmetric IRREP of the  $\mathcal{C}_{\textrm{2v}}$ point group at all times.

At this juncture, it is natural to ask how the above findings will alter if one replaces nitrogen by a more electronegative atom, oxygen. 
To explore the question further, let us analyse EFD and the difference density for furan  at three time steps as presented in Fig.~\ref{furan_stream}. 
The electronic ground state density and the NTO of the target transition in Fig.~\ref{fig1} already hint at a significant change in the charge migration dynamics.
Unlike for pyrrole, the positive and negative contours of the difference electronic densities are distributed not only around the atoms but also around the bonds,
and the latter is higher in intensity. 
They are anti-symmetric about the $y$-axis except around the oxygen atom. 
These findings are in agreement with the density difference profile observed in  Fig.~\ref{fig1}(c).

{The bond strengthening (weakening) shown by this increase (decrease) in the electronic densities around the bonds is primarily visible on the two sides of the oxygen atom.
In contrast to pyrrole, these will affect the bond strength and potentially cause bonds to break, resulting in ring opening on either side of the heteroatom.
Exciting furan with short, intense laser pulses of the form used in this work
is therefore more likely to trigger photochemical processes on longer timescales.}
Interestingly, the difference density around the oxygen is negative at all times, reflecting the larger electronegativity of oxygen compared to the carbon atoms of the ring.
Since the oxygen substitution preserves all operations of the $\mathcal{C}_{\textrm{2v}}$ point group, the density difference upon charge migration in furan
retains the same B\textsubscript{2} IRREP as for pyrrole.

From the EFD for furan displayed in Fig.~\ref{furan_stream}, the EFD reveals a swirling type of motion between the atoms of the ring.
The streamlines show a flow downward through the bonds between the oxygen and its nearest carbon atoms, and going upward to the hydrogen atoms at time steps  $\tau$ and  $2\tau$. 
The direction of the flow is completely reversed at  6$\tau$. These patterns are consistent with the NTO picture in  Fig.~\ref{fig1}(b),
which show a stronger {in-plane bonding} character and more nodal structure in the plane of the molecule for the target excitation.
Despite the apparent complexity of the flow patterns, they transform according to the totally symmetric IRREP of the  $\mathcal{C}_{\textrm{2v}}$ point group at all times.
It can be concluded that the stronger {in-plane bonding} character of the target excitation in a ring bearing a more electronegative atom does not
affect the symmetry of the charge migration.

\begin{figure*}[hbt!]
\includegraphics[width = \linewidth]{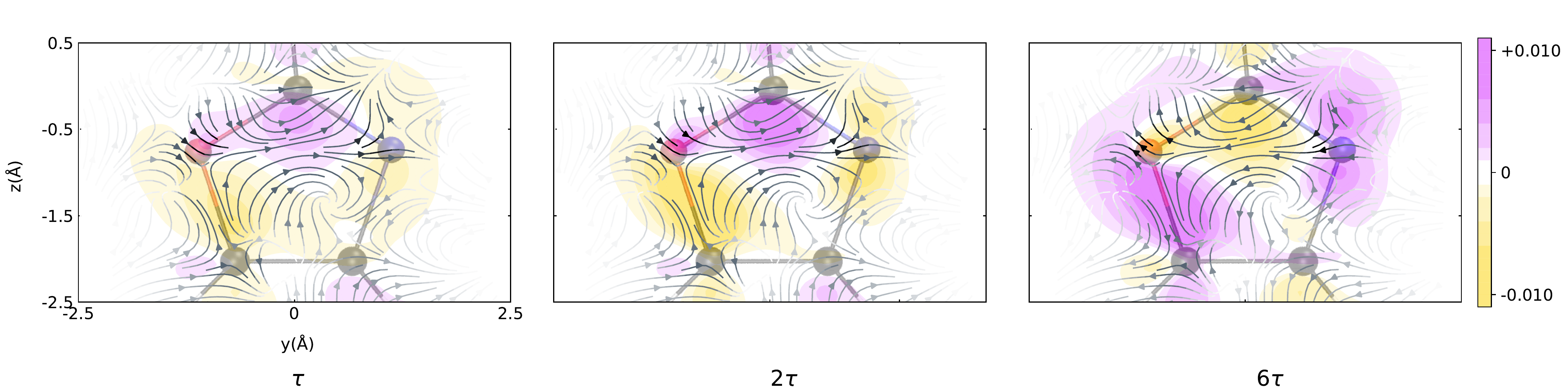}
\caption{Same as Fig.~\ref{pyrrole_stream} for oxazole with $\textrm{T} =  650$ attoseconds  as  the characteristic timescale.} 
\label{oxazole_stream}
\end{figure*}
For the charge migration in pyrrole and furan, both molecules belong to $\mathcal{C}_{\textrm{2v}}$ point group.  
The symmetry of the five-membered ring can be reduced by substitution of two carbon atoms by both nitrogen and oxygen.
To understand this further symmetry reduction, let us explore the charge migration in oxazole, which has 
$\mathcal{C}_{\textrm{s}}$ symmetry. 

As a result of the symmetry reduction, 
the charge migration in oxazole  changes drastically as two carbon atoms are replaced by one nitrogen and one oxygen atoms on the opposite sides of the pentagon ring (see Fig.~\ref{oxazole_stream}).
Due to the presence of the different atoms in the ring, 
the difference electron densities are entirely different in comparison to  pyrrole and furan, 
and do not exhibit any symmetric behaviour upon rotation with respect to the  $z$-axis or reflection in the $xz$-plane.
The regions of density increase/depletion are not specific to the bonds or atoms. 
The dumbbell structure observed at all time step in the electronic density different around nitrogen atom hints at a strong contribution
of a $p$-type orbital in the molecular plane. 
At  $\tau$,   the negative contours are present around the bonds and positive contours are situated in the space between nitrogen, oxygen and  carbon atoms.
In general, there is no strong correlation between the difference density plots during charge migration and the ones reported in Fig.~\,\ref{fig1}(c).

To have a better understanding of the charge migration mechanism in oxazole,
let us analyse the time-resolved EFD after laser excitation, as depicted in Fig.~\ref{oxazole_stream}. 
The multiple directions of the flux densities provide a picture of the charge migration that is consistent with the NTO densities depicted in Fig.~\,\ref{fig1}(b).
A swirling motion is observed, separated in bottom and top contributions that barely exchange electrons at all times.
This is intriguing, since the excitation using a $y$-polarised pulse would likely create nodal structure along this direction.
As such, a left-right separation of the charge migration should be expected, as was the case for pyrrole and furan.
From the NTOs in Fig.~\,\ref{fig1}(b), it appears that the top part of the particle density only migrates towards the hole density
located between the two heteroatoms. We attribute this feature to their electronegativity, that prefer to retain the excited electron
localised in this fragment.
On the other hand, the bottom part of the charge migration sees a transfer of the particle localised on the CO bond to the CC bond across the ring.
This part of the charge migration operates through space rather than following the bonds, in stark contrast with the more symmetrical pyrrole and furan.
{The bonds connecting to the oxygen atom experience the strongest bond strength oscillations,
whereas bonds connected to the nitrogen atom see less pronounced fluctuations. 
Much like the case of furan, it can be inferred from the electronic density and EFD snapshots
that oxazole could probably undergo further photochemical reactions.}

\section{Conclusion}  
In summary, we have investigated the role of symmetry reduction on charge migration in 
selected heterocyclic five-membered ring molecules.
Time-dependent transition electron densities and electronic flux densities were used to unravel the mechanism of charge migration
induced by selective laser pulses. 
To compare the charge migration on three molecules on equal footing, transitions with similar energetics were used to reach target excited states with {high transition intensities} and nodal structure close to the heteroatoms.
It has been found that laser-induced charge migration and the corresponding EFD are significantly different in the three molecules in spite of these constraints.
{Moreover, variations in electronic densities and EFDs reveal bond strengthening or weakening upon excitation by light,
which is useful to infer photochemical potential in these molecules.
For the laser pulses used in this work, it was inferred that pyrrole is less likely to undergo ring-opening reactions than furan and oxazole.}

The presence of nitrogen in pyrrole and oxygen in furan make the charge migration in both molecules significantly different. 
In pyrrole, a signature of charge  localisation is observed during charge migration  as the difference electronic densities are positioned  around the atoms.
In contrast, electron density depletion/increase regions are delocalised around the atoms and around the bonds during light-induced charge migration in furan,
which indicates the bond-to-atom charge transfer mechanism. The charge migration patterns observed in both cases are totally symmetric.
For oxazole, the difference electronic densities documenting the charge migration dynamics display no symmetry outside of the $yz$-plane
as a result of the lower symmetry, from $\mathcal{C}_{\textrm{2v}}$ in furan and pyrrole to $\mathcal{C}_{\textrm{s}}$. 
In this case, charge migration takes the form of  a swirling motion through space. 

We believe that our results on attosecond charge migration in heterocyclic five-membered rings will motivate
further theoretical investigations in other types of molecular systems, where studies of structure-dynamics relationships
are still few and far apart. Further, state-of-the-art emerging experimental techniques, such as attosecond transient absorption spectroscopy~\cite{kraus2015measurement, calegari2014ultrafast}, high-harmonic generation spectroscopy~\cite{folorunso2021molecular, he2022filming, dixit2018control, chandra2019experimental}, 
or time-resolved x-ray diffraction~\cite{dixit2012imaging, dixit2014theory}, to name but a few,
will provide potent probes for our findings in heterocyclic five-membered ring molecules in the coming future. 

\backmatter


\bmhead{Acknowledgments}

S. G. acknowledges Council of Scientific and Industrial Research (CSIR) 
for senior research fellowship (SRF). 
G. D. acknowledges support from Science and Engineering Research Board (SERB) India 
(Project No. MTR/2021/000138).

\section*{Declarations}
The datasets generated during and/or analysed during the current study are available from the corresponding author on reasonable request.
%


\bibliographystyle{ieeetr}
\bibliography{threemol_sorted}

\end{document}